\documentclass{article}

\usepackage{arxiv}

\usepackage[utf8]{inputenc} 
\usepackage[T1]{fontenc}    
\usepackage{hyperref}       
\usepackage{url}            
\usepackage{booktabs}       
\usepackage{amsfonts}       
\usepackage{nicefrac}       
\usepackage{microtype}      
\usepackage{lipsum}
\usepackage{pdfpages}

\usepackage[round]{natbib}

\usepackage{amsmath,amssymb,amsfonts}
\usepackage{tikz}
\usetikzlibrary{positioning, calc}
\usepackage{float}
\usepackage{graphicx}
\usepackage{subfig}
\usepackage{prodint}

\newcommand{\Prob}{\mathsf{P}}
\newcommand{\E}{\mathsf{E}}

\newcommand{\given}{\, | \,}
\newcommand{\LM}{\textrm{LM}}
\newcommand{\HH}{\textrm{H}}
\newcommand{\AJ}{\textrm{AJ}}
\newcommand{\LMAJ}{\textrm{LMAJ}}
\newcommand{\HAJ}{\textrm{HAJ}}
\DeclareMathOperator*{\cov}{cov}

\title{A hybrid landmark Aalen-Johansen estimator for transition probabilities in partially non-Markov multi-state models}

\author{
  NIKLAS MALTZAHN \\
  Oslo Centre for Biostatistics and Epidemiology \\
  Oslo University Hospital\\
  PO Box 4950 Nydalen, 0424 Oslo, Norway \\
  \texttt{niklas.maltzahn@medisin.uio.no} \\
   \And
 RUNE HOFF \\
  Oslo Centre for Biostatistics and Epidemiology \\
  Oslo University Hospital\\
  PO Box 4950 Nydalen, 0424 Oslo, Norway \\
  \AND
  ODD O. AALEN \\
  Oslo Centre for Biostatistics and Epidemiology \\
  Department of Biostatistics, University of Oslo \\
  PO Box 1122 Blindern, 0317 Oslo, Norway \\
  \And
  INGRID S. MEHLUM \\
  National Institute of Occupational Health \\
  PO Box 8149 Dep, 0033 Oslo, Norway \\
  \And
  HEIN PUTTER \\
  Department of Biomedical Data Sciences \\
  Leiden University Medical Center \\
  PO Box 9600, 2300 RC Leiden, The Netherlands \\
  \And
  JON MICHAEL GRAN \\
  Oslo Centre for Biostatistics and Epidemiology \\
  Department of Biostatistics, University of Oslo \\
  PO Box 1122 Blindern, 0317 Oslo, Norway \\
}

\begin{document}
\maketitle

\begin{abstract}
Multi-state models are increasingly being used to model complex epidemiological and clinical outcomes over time. It is common to assume that the models are Markov, but the assumption can often be unrealistic. The Markov assumption is seldomly checked and violations can lead to biased estimation for many parameters of interest. As argued by Datta and Satten (2001), the Aalen-Johansen estimator of occupation probabilities is consistent also in the non-Markov case. Putter and Spitoni (2018) exploit this fact to construct a consistent estimator of state transition probabilities, the landmark Aalen-Johansen estimator, which does not rely on the Markov assumption. A disadvantage of landmarking is data reduction, leading to a loss of power. This is problematic for “less traveled" transitions, and undesirable when such transitions indeed exhibit Markov behaviour. Using a framework of partially non-Markov multi-state models we suggest a hybrid landmark Aalen-Johansen estimator for transition probabilities. The proposed estimator is a compromise between regular Aalen-Johansen and landmark estimation, using transition specific landmarking, and can drastically improve statistical power. The methods are compared in a simulation study and in a real data application modelling individual transitions between states of sick leave, disability, education, work and unemployment. In the application, a birth cohort of \mbox{184 951} Norwegian men are followed for 14 years from the year they turn 21, using data from national registries.
\end{abstract}

\keywords{Landmarking  \and Non-Markov multi-state models  \and The Aalen-Johansen estimator  \and Transition probabilities}

\section{Introduction}
\label{sec1}

Multi-state models are increasingly being used to model complex epidemiological and clinical outcomes over time. One example is in the analysis of long-term sick leave and health related absence from work, where detailed longitudinal data on individuals are available through administrative registries (see e.g. \citet{hoff17}). Multi-state models extend traditional hazard based time-to-event models to situations with a higher, finite, number of states, each of which defines a possible competing risks situation (\citealp{hougaard99,andersen02,putter07,meira08}). In such models, the objects of interest for estimation, beside covariate effects and the transition hazards themselves, are typically occupation and transition probabilities. In a Markov multi-state model, occupation and transition probabilities can be estimated consistently as a plug-in estimate based on the estimated transition intensities using the Aalen-Johansen (AJ) estimator \citep{aalen08}. However, for non-Markov models, the AJ estimator is only consistent for occupation probabilities (\citealp{datta01, glidden02, Overgaard19, Beyersmann20}).

Several methods have been proposed for estimating transition probabilities in general semi- and non-Markov multi-state models; see for example \cite{allignol14}, \cite{titman15}, \cite{deunaalvarez15} and \cite{putter16}, who all propose methods based on subsampling. The landmark Aalen-Johansen (LMAJ) method of Putter and Spitoni is based on analysing a subset of the population who are in a specific state at a specific time point. This reference time and state is referred to as a landmark. Applying the AJ estimator to this landmark subset gives consistent estimates of transition probabilities from the landmark state at the landmark time, even for non-Markov models. A consequence of stratification to a landmark sample is that these subsets may become small, and estimation inefficient. In this paper we suggest an alternative approach, the hybrid landmark Aalen-Johansen (HAJ) estimator, for models consisting of Markov and non-Markov transitions. This hybrid estimator is based on a transition wise consideration of whether to use data from the landmark subsample or all available data in the estimation procedure. Inspired by \cite{TitmanPutter19} a type of two-sample test is suggested to select which transitions that are Markov (or close to Markov) and which are not. The resulting HAJ estimator can be seen as a compromise between the two extremes of either assuming all transitions are Markov or no transitions are Markov. As our results demonstrate, the HAJ estimator will, typically, have less bias than the AJ estimator and higher precision than the LMAJ estimator. 

The outline for the paper is as follows. In Section \ref{sec2} we define partially non-Markov multi-state processes and present the HAJ estimator. In Section \ref{sec3} we give a heuristic justification of the estimator, discuss large sample properties and  how tests of Markov behaviour following \cite{TitmanPutter19} can be used to construct the estimator. In Section \ref{sec4} we consider a simulation study comparing the HAJ estimator to the AJ and LMAJ estimators. In Section \ref{sec5} we apply the techniques to our motivating example, using data from a Norwegian birth cohort to model sickness absence and work participation over time. A discussion is found in Section \ref{sec6}. R code for implementation and reproduction of the simulation study is available on GitHub (see Supplementary Material).

\section{A hybrid landmark Aalen-Johansen estimator}
\label{sec2}

Let us consider a multi-state model $X(t)$ over a bounded time interval $[0, \tau]$, taking values in the state space $\mathcal{K} = \{1, \ldots, K\}$. Let $E \subset \mathcal{K} \times \mathcal{K}$ be the set of possible transitions of $X$. For a given state $l \in \mathcal{K}$, we are interested in the transition probabilities from $l$ to each of the states in $\mathcal{K}$, given by
\begin{equation}
\mathbf{P}_l(s, t) := \left( P_{l1}(s,t). \ldots, P_{lK}(s,t) \right)^\top, \label{intro1}
\end{equation}
where $P_{lk}(s,t) := \Prob(X(t) = k \given X(s) = l)$ and $k \in \{1,...,K\}$. The problem of interest is to produce a consistent estimator of \eqref{intro1}.

Note that the theory that follows also is valid for more than one landmark state $l$, so that $l$ in practice can be a set of states. However, to ease notation, we focus on the most common scenario where $l$ is one particular state in the state space $\mathcal{K}$.

\subsection{The Aalen-Johansen estimator}

When the multi-state process is Markov, a consistent estimator of these transition probabilities is provided by the Aalen-Johansen estimator \citep{AJ78}. For this, consider $n$ i.i.d. realisations $X_i(t)$ of $X(t)$, where for subject $i$ we define the at risk process for state $j$ as $Y_j^{(i)}(t) = 1\{ X_i(t-) = j\}$ and the transition counting process for transition $j \to k \in E$ as $N_{jk}^{(i)}(s,t) = \sum_{u \in (s,t]} 1\{X_i(u-) = j, X_i(u) = k\}$. Then, define the aggregated at risk and transition counting processes as
\begin{align*}
    \overline{Y}_{j}(t) := \sum_{i = 1}^{n} 1\{X_i(t-) = j\} 
    \quad \text{and} \quad
    \overline{N}_{jk}(t) := \sum_{i = 1}^{n} \sum_{u \in (s,t]} 1\{X_i(u-) = j, X_i(u) = k\},
\end{align*}
and let $\overline{\mathbf{Y}}(t) := (\overline{Y}_{1}(t), \ldots, \overline{Y}_{K}(t))$, and $\overline{Y}_{\bullet}(t) = \sum_{j=1}^K \overline{Y}_j(t)$ be the total number of subjects at risk at time $t$. For $J_{j}(t) := 1\{\overline{Y}_j(t) > 0\}$ let
\begin{align}
\widehat{\Lambda}_{jk}(t) := \int_{0}^{t} \frac{J_{j}(u) \mathrm{d}\overline{N}_{jk}(u)}{\overline{Y}_{j}(u)} \label{NA1}
\end{align}
be the Nelson-Aalen estimator of the transition rates and $\widehat{\mathbf{\Lambda}}(t)$ a matrix with $(j,k)$th element equal to $\widehat{\Lambda}_{jk}(t)$ and diagonal elements $\widehat{\Lambda}_{jj}(t) = - \sum_{k \not= j} \widehat{\Lambda}_{jk}(t)$. The Aalen-Johansen (AJ) estimator of the transition probability matrix $\mathbf{P}(s, t)$, with elements $P_{jk}(s,t)$, is then given by
\begin{align*}
\widehat{\mathbf{P}}^{\AJ}(s,t) &:= \prod_{u \in (s,t]} \left(\mathbf{I} + \Delta\widehat{\mathbf{\Lambda}}(u) \right).
\end{align*}
Estimated state occupation probabilities at time $t$ may be obtained by 
\begin{equation}\label{eq:stateocc}
    \widehat{\mathbf{\pi}}^{\AJ}(t) = \widehat{\mathbf{\pi}}(0) \widehat{\mathbf{P}}^{\AJ}(0,t),
\end{equation}
where $\widehat{\mathbf{\pi}}(0)$ is the row vector of empirical state occupation probabilities at $t=0$, given by $\widehat{\pi}_j(0) := \overline{Y}_j(0+) / \overline{Y}_{\bullet}(0+)$. Here, the $+$ in $(t+)$ means that the numbers observed to be in state $j$ \emph{at} time $t$, rather than just before time $t$ are to be taken. \citet{datta01} argued that the estimated state occupation probabilities in~\eqref{eq:stateocc} are consistent, even if the multi-state model is non-Markov. The Aalen-Johansen (AJ) estimator of the transition probabilities $\mathbf{P}_l(s, t)$ from \eqref{intro1} is then given by
\begin{align*}
\widehat{\mathbf{P}}_{l}^{\AJ}(s,t) &:= e_{l} \prod_{u \in (s,t]} \left(\mathbf{I} + \Delta\widehat{\mathbf{\Lambda}}(u) \right),     
\end{align*}
where $e_{l}$ is a vector with the $l$th element equal to 1, and all other elements 0.

\subsection{The landmark Aalen-Johansen estimator}

Building on the results of \cite{datta01}, \cite{putter16} defined the landmark Aalen-Johansen (LMAJ) estimator of transition probabilities. This estimator uses the landmark population $\{i: X_i(s) = l\}$ for estimation of the transition intensities. In what follows we consider the landmark time $s$ and landmark state $l$, on which we condition, as fixed. Let us suppress the dependence on $s$ and $l$ in the notation when defining the landmark counting process and at risk process as $Y_j^{(i, \LM)}(t) := 1\{X_i(s) = l\}Y_j^{(i)}(t)$ and $N_{jk}^{(i, \LM)}(t) := 1\{X_i(s) = l\}N_{jk}^{(i)}(s,t)$. We define the aggregated at risk and counting processes based on the landmark population as
\begin{align*}
    \overline{Y}_{j}^{(\LM)}(t) := \sum_{i = 1}^{n} Y_j^{(i, \LM)}(t) \quad \text{and} \quad \overline{N}_{jk}^{(\LM)}(t) := \sum_{i = 1}^{n} N_{jk}^{(i, \LM)}(s,t).
\end{align*}
Let $\overline{\mathbf{Y}}^{(\LM)}(t) := (\overline{Y}_{1}^{(\LM)}(t), \ldots, \overline{Y}_{K}^{(\LM)}(t))$ and $\overline{Y}_{\bullet}^{(\LM)}(t) := \sum_{j = 1}^{K} \overline{Y}_{j}^{(\LM)}(t)$. 

For $J_{j}^{(\LM)}(t) := 1\{\overline{Y}_{j}^{(\LM)}(t) > 0\}$ define the landmark Nelson-Aalen estimator of the transition rates as
\begin{align}
\widehat{\Lambda}_{jk}^{(\LM)}(t) := \int_{s}^{t} \frac{J_{j}^{(\LM)}(u) \mathrm{d}\overline{N}_{jk}^{(\LM)}(u)}{\overline{Y}_{j}^{(\LM)}(u)}, \label{NALMAJ1}
\end{align}
and let $\widehat{\mathbf{\Lambda}}^{(\LM)}(t)$ be the matrix with $(j,k)$th element $\widehat{\Lambda}_{jk}^{(\LM)}(t)$ and diagonal element $\widehat{\Lambda}_{jj}^{(\LM)}(t) = - \sum_{k \not= j} \widehat{\Lambda}_{jk}^{(\LM)}(t)$. Then the landmark Aalen-Johansen (LMAJ) estimator of \eqref{intro1} presented by \cite{putter16} is given by
\begin{align*}
\widehat{\mathbf{P}}_{l}^{\LMAJ}(s,t) &:= e_{l} \prod_{u \in (s,t]} \left(\mathbf{I} + \Delta\widehat{\mathbf{\Lambda}}^{(\LM)}(u) \right).     
\end{align*}

\subsection{The hybrid landmark Aalen-Johansen estimator}

An undesirable feature of landmark subsampling is a reduction of the number of individuals at risk used for estimation, for all transitions, including Markov transitions if such exist. An easy way of improving the estimation is by plugging in Nelson-Aalen estimates based on the landmark sample in the Aalen-Johansen (AJ) estimator only for transitions which causes non-Markov behaviour and this is the main idea behind what we will refer to as the hybrid Aalen-Johansen (HAJ) estimator. For such an estimation procedure, we are thus in particular interested in detecting how violations from the Markov assumption are attributable to specific transitions. If this is only a subset of the full set of transitions $E$, we refer to the multi-state processes as \emph{partially non-Markov} and denote the set of non-Markov transitions as $A \subset E$. 

The HAJ estimator considers as estimators of the transition rates
\begin{align}
\widehat{\Lambda}_{jk}^{(\HH)}(t) := 
    \left\{
          \begin{array}{ll}
            \widehat{\Lambda}_{jk}(t), & \hbox{$jk \not\in A$;} \\
            \widehat{\Lambda}_{jk}^{(\LM)}(t), & \hbox{$jk \in A$.}
          \end{array}
        \right.
    \label{NAHAJ1}
\end{align}
Define $\widehat{\mathbf{\Lambda}}^{(\HH)}(t)$ to be the matrix with $(j,k)$th element $\widehat{\Lambda}_{jk}^{(\HH)}(t)$ and diagonal element $\widehat{\Lambda}_{jj}^{(\HH)}(t) = - \sum_{k \not= j} \widehat{\Lambda}_{jk}^{(\HH)}(t)$. Now, the HAJ estimator of \eqref{intro1} is
\begin{align*}
\mathbf{\widehat{P}}_{l}^{\HAJ}(s,t) &:= e_l \prod_{u \in (s,t]} (\mathbf{I} + \Delta\widehat{\mathbf{\Lambda}}^{(\HH)}(u)).  
\end{align*}

Observe that for $A = \emptyset$ we get the classical AJ estimator, while for $A = E$ we get the LMAJ estimator. If we assume Markov behaviour only for certain transitions, this means in practice that we can implement the HAJ estimator by removing individuals that are not in the landmark state at the landmark time point from the specific risk sets used for estimating intensities for the non-Markov transitions. Applying the AJ estimator to such a reduced dataset from the landmark time point and onward will produce the HAJ estimate. As already mentioned, the LMAJ estimator is expensive; meaning that data reduction reduces precision (increases variance). The HAJ estimator will guarantee equal or better precision, at the possible expense of introducing bias. In Section~\ref{sec4} we will study how these opposing effects balance out.

\section{Justification of the HAJ estimator}
\label{sec3}

\subsection{Product limits and transition probabilities}
\label{sub31}

Estimation of transition probabilities in multi-state models relies on a special relation between conditional probabilities and cumulative hazard functions. The relation is the multi-state version of the argument for the Kaplan-Meier estimator in classical time-to-event models. For Markov multi-state models the result is due to \cite{gill90} and says that if $\mathbf{P}(s, t)$ is the transition probability matrix of a Markov multi-state model and $\mathbf{\Lambda}$ the cumulative hazard rate matrix, then we have 
\begin{align}
\mathbf{P}(s, t) = \lim \prod_{m = 1}^{M} \mathbf{P}(t_{m-1},t_{m}) = \lim \prod_{m = 1}^{M} \left(I + \mathbf{\Lambda}(t_{m})  - \mathbf{\Lambda}(t_{m-1}) \right), 
\label{refine0}
\end{align}
where the limits are taken over refinements of $(s,t]$, and $t_0=s$ and $t_M=t$.  As a consequence, the product integral, when considered as a functional 
\begin{align}
\mathbf{\Lambda} \to  \Prodi_{u \in (s, t]}(I + \mathrm{d}\mathbf{\Lambda}(u)), \label{functional}
\end{align}
is a convenient construct for producing plug-in estimators of probabilities in multi-state models based on estimators of the cumulative hazard matrix. Utilizing the relation of the product integral above to Duhamel's equation (see e.g. \citet[Chapter 2]{andersen93}), two ways of deriving large sample properties of plug-in estimators are obtained; one based on the functional delta method and empirical process theory exploiting smoothness properties of \eqref{functional} and one based on martingale theory (see e.g. \citet[ p. 320]{andersen93}). In the non-Markov case two problems arise. The main problem is that the result of Gill and Johansen establishing \eqref{refine0} no longer holds and thus the expression of probabilities through the product integral requires a new argument. This is resolved by \cite{Overgaard19} in a similar spirit as that of \cite{gill90}. In the non-Markov case the transition rates are no longer compensators of the transition counting processes and thus the martingale property no longer holds. This makes it unclear if the argument by \cite{datta01} is in fact valid. A concern also expressed and remedied by \cite{Beyersmann20}. However, consistency of the Aalen-Johansen estimator of state occupation probabilities still holds and a proof is given by \cite{Overgaard19}. The same is true regarding the landmark estimator and a proof can be found in \cite{Beyersmann20}. We do not go into details in terms of large sample properties of the HAJ estimator. Instead we will give a heuristic justification of the estimator and rely on the results of \cite{Overgaard19} and \cite{Beyersmann20}. 

In order to justify the HAJ estimator we first consider the LMAJ estimator. As earlier, consider a fixed landmark time point $s$ and landmark state $l$  and define $\mathbf{P}_l(t, u) = \Bigl( P_{l,jk}(t, u) \Bigr)$ to be the matrix of transition probabilities in \emph{the landmark population}, with
\begin{equation*}
P_{l,jk}(t, u) := \Prob(X(u) = k \given  X(t) = j, X(s) = l).
\end{equation*}
By construction, and regardless of the Markov assumption, we have for $s \leq t$
\begin{align*}
\mathbf{P}_{l}(s,t) = e_l \mathbf{P}_l(t_{0},t_{1}) \cdots \mathbf{P}_l(t_{M-1},t_{M}),
\end{align*}
for $s = t_{0} < t_{1} < \cdots < t_{M} = t$. For a sufficiently fine partition of the interval $(s,t]$ consider the approximation $\mathbf{P}_l(t_{l-1}, t_{l}) \approx I + \mathbf{\Lambda}^{(\LM)}(t_{l})  - \mathbf{\Lambda}^{(\LM)}(t_{l-1})$ where $\Lambda_{jk}^{(\LM)}(\mathrm{d}t)$ is the transition rate of the landmark population, i.e. 
\begin{align*}
\Lambda_{jk}^{(\LM)}(\mathrm{d}t) = \E\left[ \mathrm{d} N_{jk}^{(\LM)}(t) \given  Y_{1j}^{(\LM)}(t) = 1 \right].
 \end{align*}
The desired result is then to achieve the equality
\begin{align}
\lim \prod_{m = 1}^{M} \mathbf{P}_l(t_{m-1},t_{m}) = \lim \prod_{m = 1}^{M} \left(I + \mathbf{\Lambda}^{(\LM)}(t_{m})  - \mathbf{\Lambda}^{(\LM)}(t_{m-1}) \right), \label{refine}
\end{align}
where the limits are taken over refinements of $(s,t]$. For the Markov case, this is derived by \cite{gill90}, and for the non-Markov case, it is derived by \cite{Overgaard19}. Hence also in the non-Markov case an estimator of the transition probability $\mathbf{P}_{l}(s,t)$ may be obtained when we have a consistent estimator of the right hand side of \eqref{refine}. 

\subsection{The LMAJ estimator}

Assuming no censoring one can, as pointed out by \cite{aalen01} and alternatively by \citet[p. 296]{andersen93}, derive the landmark estimator of $\mathbf{P}_{l}(s,t)$ from a bookkeeping argument. In order to do so let 
\begin{align*}
\widehat{\mathbf{P}}^{(\LM)}(s,t) &= \prod_{u \in (s, t]}(I + \Delta\widehat{\mathbf{\Lambda}}^{(\LM)}(u)).
\end{align*}
A natural way of estimating the transition probabilities $\mathbf{P}_{l}(s,t)$ for specific time points $s$ and $t$ is to consider the fraction of empirical means $\mathbf{\overline{Y}}^{(\LM)}(t) / \overline{Y}_{\bullet}^{(\LM)}(s)$. The number of individuals from the landmark sample in state $i$ just after time $s$, $\overline{Y}_{j}^{(\LM)}(s + \Delta t)$, may be expressed as those who where in state j at time $s$ plus the net arrivals in the small time frame $\Delta t$. That is
\begin{align*}
\overline{Y}_{j}^{(\LM)}(s + \Delta t) &= \overline{Y}_{j}^{(\LM)}(s) + \sum_{k \neq j}  (\overline{N}_{kj}^{(\LM)}(s + \Delta t) - \overline{N}_{kj}^{(\LM)}(s)) \ - \\
& \quad \sum_{k \neq j}  (\overline{N}_{jk}^{(\LM)}(s + \Delta t) - \overline{N}_{jk}^{(\LM)}(s)) \\
&\to \overline{Y}_{j}^{(\LM)}(s) [I + \Delta\widehat{\mathbf{\Lambda}}^{(\LM)}(s +)]_{j}
\end{align*}
for $\Delta t$ $\to 0$,
where $[I + \Delta \widehat{\mathbf{\Lambda}}^{(\LM)}(s +)]_{j}$ denotes the $j$th column of $I + \Delta \widehat{\mathbf{\Lambda}}^{(\LM)}(s +)$. From this observation one obtains the following algebraic relation  
\begin{align}
\frac{ \mathbf{\overline{Y}}^{(\LM)}(t) }{\overline{Y}_{\bullet}^{(\LM)}(s)} &= \frac{\mathbf{\overline{Y}}^{(\LM)}(s) \widehat{\mathbf{P}}^{(\LM)}(s,t) }{\overline{Y}_{\bullet}^{(\LM)}(s)}  = e_{l} \widehat{\mathbf{P}}^{(\LM)}(s,t)   = \mathbf{\widehat{P}}_{l}^{\LMAJ}(s,t). \label{LMAJ2}
\end{align}
A different conclusion to draw from \eqref{LMAJ2} is that if the equality 
\begin{align*}
    \frac{ \mathbf{\overline{Y}}^{(\LM)}(t) }{\overline{Y}_{\bullet}^{(\LM)}(s)} &= \hat{\pi}^{(\LM)}(s) \widehat{\mathbf{P}}^{(\LM)}(s,t)
\end{align*}
should be satisfied when using the plug-in estimator based on the product integral, then the cumulative transition hazards matrix has to be based on Nelson-Aalen estimates obtained from the landmark data.

\subsection{The HAJ estimator}

If $X_{1}, X_{2}, \ldots$ are partially non-Markov, then there is potential gain in terms of power and variance when using the HAJ estimator. Note that 
\begin{align}
\E\left[ \mathrm{d} N_{jk}^{(\LM)}(t) \given  Y_{j}^{(\LM)}(t) = 1 \right] = \E\left[ \mathrm{d} N_{jk}(t) \given  Y_{j}(t) = 1, X(s) = l \right], \label{LMNA}
\end{align}
which for Markov transitions $(jk) \in A$ implies that $\mathrm{d}\Lambda_{jk}^{(\LM)}(t) = \mathrm{d}\Lambda_{jk}(t)$. Likewise if we let $u \to \mathbf{\Lambda}^{(\HH)}(u)$ be the hybrid cumulative hazard matrix  based on $\Lambda_{jk}$ for $(jk) \in A$ and $\Lambda_{jk}^{(\LM)}$, for $(jk) \in A^{C}$, we have $\mathbf{\Lambda}^{(\LM)} = \mathbf{\Lambda}^{(\HH)}$. Thus, when \eqref{refine} holds, the same holds in the partially non-Markov setting using $\mathbf{\Lambda}^{(\HH)}$. Given the consistency of the LMAJ estimator, e.g. following from \cite{putter16} and \cite{Overgaard19}, consistency will also hold for the HAJ estimator.

As pointed out by \cite{TitmanPutter19} from \eqref{LMNA} it is clear that a test of the Markov assumption for a specific jump transition process from state $j$ to state $k$ can be obtained by comparing intensities based on disjoint landmark states. One can use a two-sample test of the hypothesis
\begin{align*}
H_{0}:\lambda_{jk}^{l_1}(s,t) = \lambda_{jk}^{l_2}(s,t) \text{ on } (s, \tau].
\end{align*} 
Note that the two disjoint landmark states (denoted with superscript) ensure independence between samples in the estimation procedure. Typically, $l_1$ would be the landmark state of main interest and $l_2$ the set of all other remaining possible states. In the application and the simulation study we use two different tests as selection criteria for determining Markov and non-Markov behaviour for specific transitions. Denote the log-rank test statistic for transition $i \to j$ from landmark time point $s$ by $\mathfrak{X}_{s}$. This test statistic is the basis for a test referred to as the \textbf{point test}. Since the above test statistic is dependent on $s$,  \cite{TitmanPutter19} suggest a more global test of the Markov assumption for transition $i \to j$ based on $\mathfrak{X} := \max_{i} \mathfrak{X}_{s_{i}}$, over a suitable grid $s_{1}, \ldots, s_{k}$. We refer to this as the \textbf{grid test}. For further discussion of the use of two sample tests to identify non-Markov transition see Web Appendix D.

\subsection{Censoring, covariates and variance}

For simplicity, we have so far not considered censoring. Censoring is not used in the simulation experiment and is not a major issue in the practical application. In time-to-event analysis censoring is generally the rule rather than the exception. Note also that stronger censoring assumptions typically are needed in the setting of multi-state models, compared to the regular survival setting (see e.g. \citet[p. 123]{aalen08}). \cite{Overgaard19} derives the result \eqref{refine0}, which would extend to \eqref{refine}, based on a form of independent censoring coined "the status independent observation assumption" and \cite{glidden02} considers right censoring with strong independence assumptions requiring censoring to be independent of states. \cite{datta02} consider an IPCW version of the AJ estimator of state occupation probabilities under dependent censoring. This weighted estimator was empirically investigated by \cite{gunnes07} and showed reasonable results for non-Markov behaviour induced by a joint frailty at baseline and various dependent censoring regimes. We have not considered covariate based hazard models in our formal treatment of the HAJ estimator but the results are expected to generalize to classical covariate models e.g. Cox or additive hazard regression due to the smoothness property of \eqref{functional}. See e.g. \cite{hoff19} for a discussion of covariate based models for the transition rates in relation to the LMAJ estimator. Regarding variance estimates for the LMAJ and HAJ estimator we recommend bootstrapping. Many of the classical variance estimators for the AJ estimator rely on the transition rates being the compensators of the transition counting processes. This is true under the Markov assumption but not if we relax the Markov assumption. We therefore expect additional variation and a formal argument (similar to that of \cite{Beyersmann20}) is included in Web Appendix A, where it is made explicit how the non-Markov behaviour induces additional variation for the Nelson-Aalen estimator. For the first simulation experiment we study the empirical coverage of the Greenwood type estimator for confidence intervals and in the practical application we compare such confidence intervals to bootstrapped confidence intervals. The results are included in Web Appendix B.2 and C.2. Although the results suggest only minor deviations of the Greenwood type estimator, this is not expected to hold in general.

\section{A simulation study}
\label{sec4}

\begin{figure}[H]
\begin{center}
\includegraphics[width=0.7\textwidth]{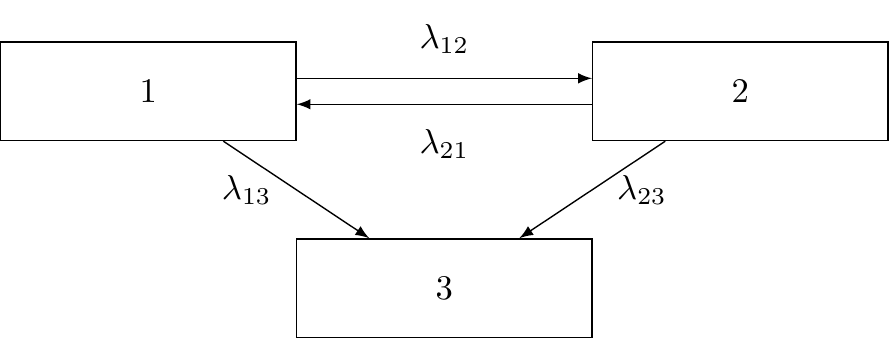}
\caption{An illness-death model with recovery, where, for example, state 1 correspond to employment, state 2 to sick leave and state 3 to permanent disability.}
    \label{ID_mod}
  \end{center}
\end{figure}

A central feature of our motivating data application on sickness absence and work participation is recurrent periods of sick leave. Since previous individual health history is very likely to impact future events of sick-leave, we expect to see non-Markov behaviour for various transitions in our model. Furthermore, we expect a considerable amount of individual heterogeneity in a number of transitions due to for example variations in socioeconomic status, educational and professional backgrounds. Motivated by these problems we will focus on a multi-state simulation experiment with recurrent events and transition intensities subject to frailty effects. The smallest relevant multi-state model for such an investigation is the illness-death model with recovery depicted in Figure \ref{ID_mod}. However, the appropriate analogy to our application will not be illness and death. Rather we will think of state 1 as employed, state 2 as on sick leave and state 3 as permanent disability.

We consider two types of experiments using non-Markov models over the time interval $[0, \tau]$, with $\tau = 1000$. In both experiments the jump transition intensities given by
\begin{align*}
\lambda_{jk} = V_{jk} \alpha_{jk}, \quad \text{for } jk \in \{(1,2), (1,3), (2,1), (2,3)\},
\end{align*}
where $(\alpha_{12}, \alpha_{13}, \alpha_{21}, \alpha_{23}) = (0.12, 0.03, 0.15, 0.1)$ and $V_{jk}$ are individual frailties. The simulation experiments were performed 1000 times and each experiment had a total sample size of 1000 individuals. The "true" transition probabilities were calculated from a separate simulation experiment as the mean over 1000 repetitions of the LMAJ estimator applied to each simulated experiment, each of which also with a sample size of 1000 individuals. Note however that the LMAJ estimates here are based on landmark subsamples which will have somewhat lower sample size than the total sample size of 1000. The asymptotic distribution from wild bootstrapping is based on 500 bootstrap samples using standardized compensated Poisson processes. We hope to find that the HAJ estimator is a useful intermediary between the AJ and the LMAJ estimator. To investigate this claim we consider two experiments; one focused on how large frailty effects need to be for the HAJ estimator to be preferable to the AJ estimator and one focused on when the non-Markov behaviour is significant enough for the HAJ estimator to compete with the LMAJ estimator. The two experiments are:

\begin{enumerate}
    \item $V_{12} = V_{13} = V_{23} = 1$ and $V_{21}$ is gamma distributed with mean 1 and variance $\sigma^{2} \in \{0, 0.4, 1.2, 2\}$, ranging from no frailty (i.e.~Markov) to heavily right-skewed frailty;
    \item $V = (V_{1}, V_{2}, V_{3}, V_{4})$ is log-normal distributed with mean 1 and covariance matrix 
\begin{align*}
\Sigma \approx \begin{pmatrix}
0.80 & 0.57 & -0.35 & 0.37 \\
0.57 & 0.42 & -0.12 & 0.19 \\
-0.35 & -0.12 & 0.96 & -0.63 \\
0.37 & 0.19 & -0.63 & 0.45
\end{pmatrix}.
\end{align*}
Then, $W = \log(V)$ is normally distributed with $EW_{j} = -\Sigma_{jj}/2$ and $\cov(W_{j}, W_{k}) = \log(1 +\Sigma_{jk})$.
\end{enumerate}

As mentioned, the HAJ estimator can be seen as a compromise between two extremes and the two experiments investigate to what extend such a compromise is useful. In other words do we need the HAJ estimator at all and if so how much do we gain by using it? 

In the first experiment the question of interest is how large the frailty variance should be in order to detect a change in the transition probabilities. In particular we are interested in at what point the HAJ estimator starts to outperform the standard AJ estimator. The second experiment investigates how the HAJ estimator performs compared to the LMAJ estimator under clear non-Markovian conditions induced by large correlated frailties (hence the choice of $\Sigma$). We evaluate the performance of the estimators using two performance measures. We consider point-wise (in time) empirical bias variance estimates (see Figure \ref{bias_var}) and mean residual squared error (MRSE) (see Figure \ref{MRSE_mod1} and \ref{MRSE_mod2}). MRSE is measured by the $L_{2}$ distance $\Vert f - g \Vert^{2} = \int_{s}^{\tau} (f(t) - g(t))^2 \mathrm{d}t$ between estimates of $t \to P_{lk}(s,t)$ and the (simulated) true transitions probability, with $\tau = 1000$. Here $s$ is landmark grid times and the $L_{2}$-distance is calculated as a Riemann sum over all jump times. All empirical performance measures are produced based on 1000 simulations of each of the model specifications above. In both experiments the HAJ estimator is constructed using the grid test.

\subsection{Experiment 1}

\begin{figure}[H]\centering
\textbf{MRSE}\par\medskip
\subfloat{\includegraphics[width=.65\linewidth]{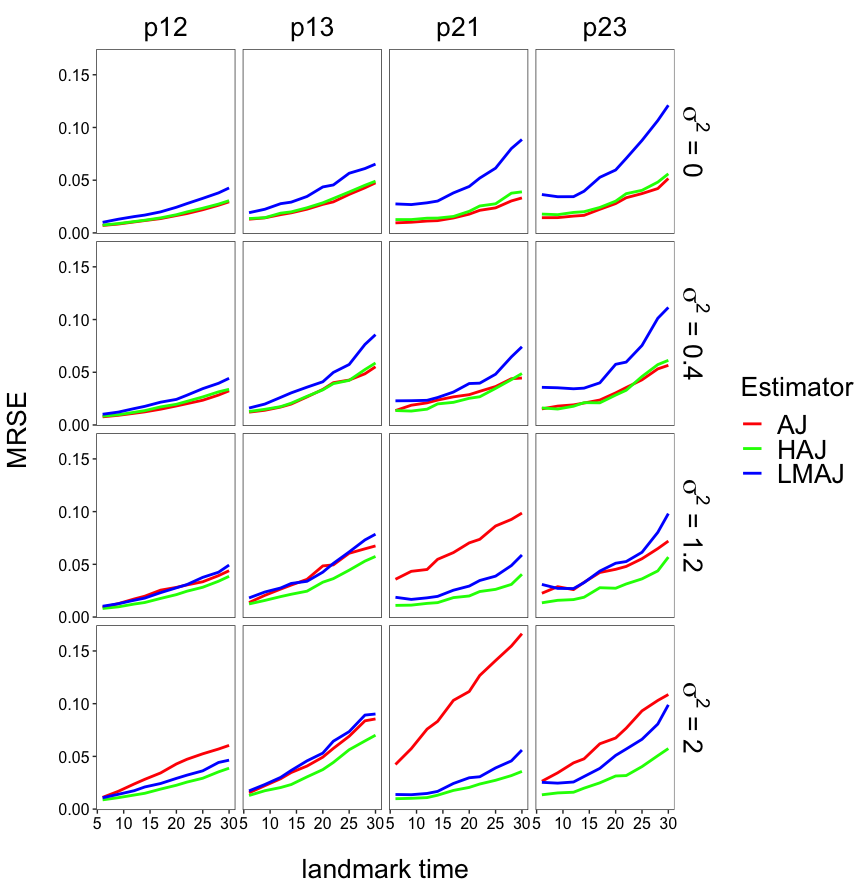}}
\caption{Mean residual squared error for the transition probabilities from state $i$ to state $j$ as a function of landmark time points. All numbers are based on 1000 samples where each sample has a size of 1000 individuals. The landmark grid is $\{6, 9, 12, 14, 17, 20, 22, 25, 28, 30\}$.}
\label{MRSE_mod1}
\end{figure}

From Figure \ref{MRSE_mod1} we see that, for non-zero frailty variance, the HAJ estimator performs at least as good as or better than the AJ estimator and better than the LMAJ estimator. In other words, the interpretation of the HAJ estimator as an intermediary between the AJ estimator and the LMAJ estimator seems to hold. In the Markov case, i.e. $\sigma^{2} = 0$, we see almost no performance difference between the HAJ and the AJ estimators. Due to the in-built error from testing the Markov assumption we generally expect better performance of the AJ estimator over the HAJ estimator for Markov models, in particular for models with many transitions. For $\sigma^{2} = 0.4$, the LMAJ estimator performs worse than the AJ estimator. The HAJ estimator is comparable to AJ for $\sigma^2 = 0.4$, but starts to outperform AJ for larger values of $\sigma^2$. In this experiment, HAJ always performed better than LMAJ. In other words, Figure \ref{MRSE_mod1} suggests that if a perturbation of the transition intensity is sufficiently large, so as to induce significant non-Markov behaviour for the transition probabilities, then the HAJ estimator outperforms the AJ and the LMAJ  estimator. However, as we shall see in the next experiment, this conclusion has its limitations. In the experiment all transitions are tested and results from the point tests and grid tests are included in Web Appendix B.1.1. 

\begin{figure}[H]\centering
\textbf{Bias-variance estimates from the simulated data for the transition probability from state 2 to state 1}\par\medskip
\begin{tabular}{c}
         \subfloat{\includegraphics[width=.90\linewidth]{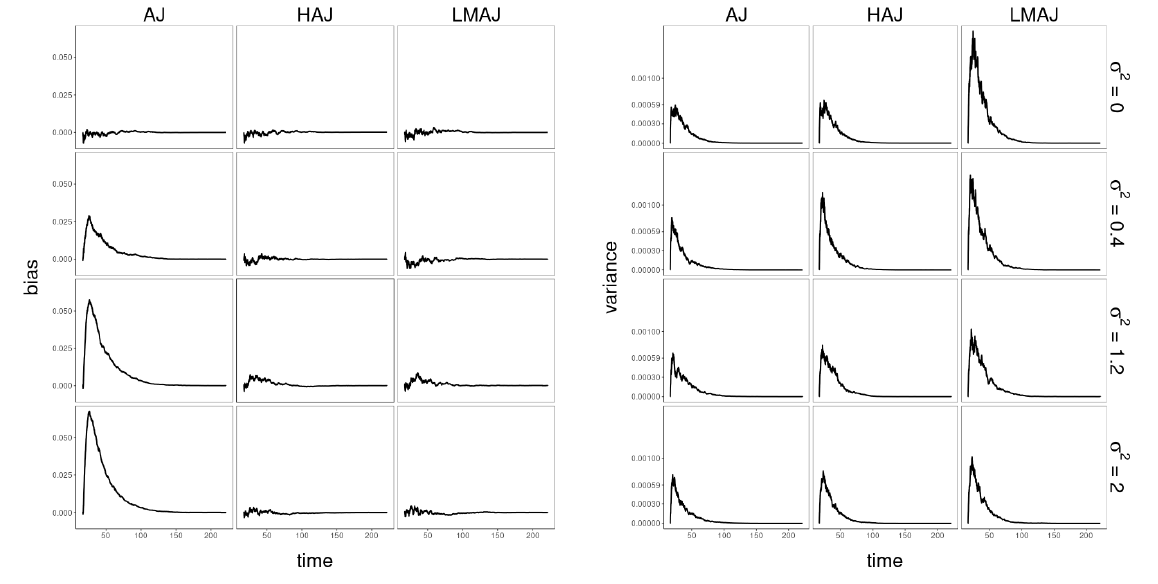}}
\end{tabular}
\caption{Bias and variance estimates for the transition probability from state 2 to state 1 based on the AJ , HAJ  and LMAJ estimator respectively. All estimates are computed from landmark time $s = 17$. Mean bias and variance estimates are based on 1000 samples, where each sample has a size of 1000 individuals.}
\label{bias_var}
\end{figure}

Regarding the bias-variance trade-off, we see from Figure \ref{bias_var} that the AJ estimator overestimates the transition probability $P_{21}(s, t)$, whereas the LMAJ and the HAJ estimator are close to the true transition probability. We also see that the HAJ estimator has smaller variance than the LMAJ estimator. This is exactly what we would expect from the HAJ estimator. Through the selection method it accounts for the partly non-Markov and partly Markov behaviour, resulting in smaller bias than the AJ estimator and less variance than the LMAJ estimator. 

\subsection{Experiment 2}

\begin{figure}[H]\centering
\begin{tabular}{c c}
        \textbf{MRSE} \\
        \subfloat{\includegraphics[width=.6\linewidth]{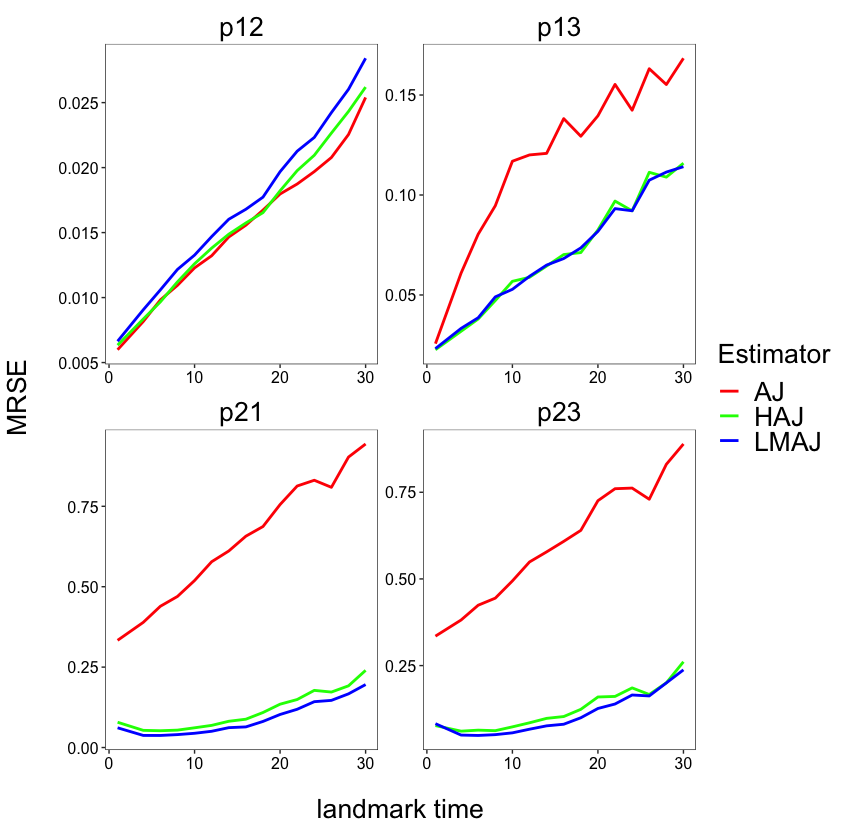}}
\end{tabular}
\caption{Mean residual squared error for the transition probability estimates from state $j$ to state $k$. The landmark grid is $s = 1, 4, 6, 8, 10, 12, 14, 16, 18, 20, 22, 24, 26, 28, 30$.}
\label{MRSE_mod2}
\end{figure}

From Figure \ref{MRSE_mod2} we see that for the estimated transition probability from state 2 to 1 and state 2 to 3 the HAJ estimator performs slightly worse than the LMAJ estimator. This is expected when the non-Markov behaviour is strong enough, simply because the selection procedure for the HAJ estimator will make the wrong choice in some percentage of cases based on the significance level. In this way the significance level of the test used in the selection procedure for the HAJ estimator becomes a tuning parameter which could be optimised. The optimal choice would be model dependent and depends on sample size and on how sensitive changes in the transition probabilities are to changes in the hazards. In the experiment all transitions are tested and results from the point tests and grid tests are included in Web Appendix B.1.2.    

\section{An application to Norwegian registry data on sick leave, disability and work participation}
\label{sec5}

\begin{figure}[H]
\begin{center}
\includegraphics[width=0.6\textwidth]{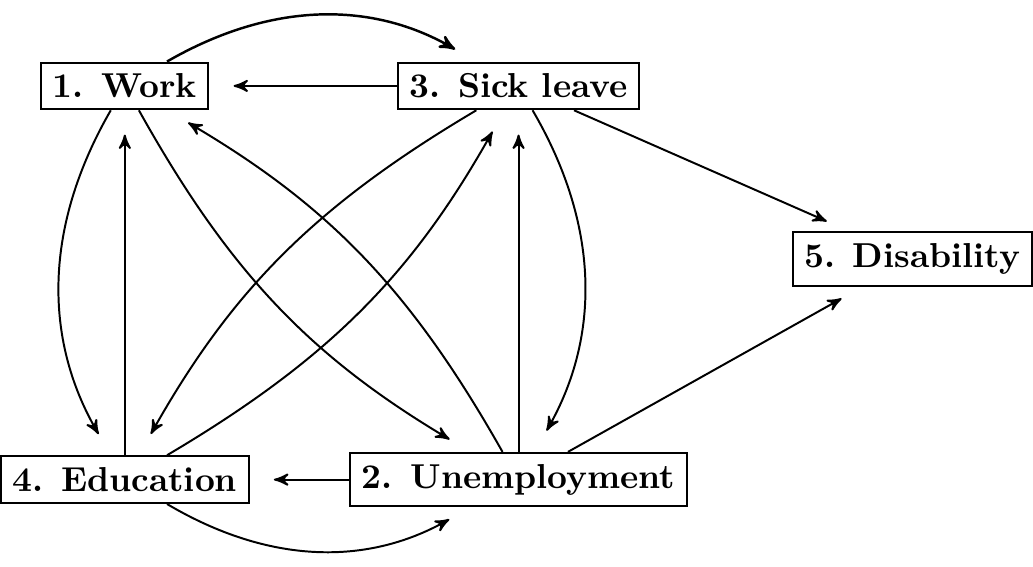}
\caption{A multi-state model for work, education and health-related absence from work.} 
    \label{fig:msm1} 
  \end{center}
\end{figure}

To illustrate the HAJ estimator and compare it to LMAJ and AJ, we consider a multi-state model with five states related to work participation. The proposed model is shown in Figure \ref{fig:msm1} and consists of the following states: (1) work, (2) unemployment, (3) sick leave, (4) education (above high school) and (5) disability, where disability is an absorbing state. Individual multi-state histories through these states are constructed using data from various Norwegian national registries with data on employment, education and welfare benefits. For more details on the data material and source registries, see \cite{hoff19}. In the dataset, information is available for the period 1992 -- 2011 for all Norwegian males born between 1971 and 1976 (n = 184 951). Additionally, several individual covariates, on socioeconomic background, health, acquired education levels and results from military conscript examination, were available. We included individuals in the study from the 1st of July the year they turned 21 (1992-1997) and observed them for 14.5 years, until 31st of December (2006-2011). The time scale used in the model is days since inclusion, so that transition times are matched on age and season, but not year.

One should realise that there are several potential violations to the Markov assumption for the model represented by Figure \ref{fig:msm1}. For example, it is plausible that individuals who have been working for a longer period are in more stable positions, further prolonging their stay in employment. Another example is individuals on long-term sick leave due to serious diseases, who may have lower probabilities for returning to work than people with minor illnesses. There are also laws and regulations that limit the possible duration of stays in states that are based on welfare benefits. Based on these circumstances we consider two examples comparing the AJ, LMAJ and HAJ estimators. In particular, we look at transitions from sick leave at time $t = 100$ (100 days since inclusion) and transitions from unemployment at time $t = 3000$ (3000 days since inclusion). 

\subsection{Example 1: Transitions from sick leave at $t = 100$}

In our first example, we calculate transition probabilities from being on sick leave at day 100. In this particular analyses, the full dataset is reduced from 184 951 individuals to 23 288 by fixing a number of covariates. Specifically, we consider high school completers attending general education (non-vocational) that scored between 7 and 9 (i.e. high scores) on the cognitive test during military conscript examinations (scores range between 1-9 where 9 is considered the best). The landmark subset then consists of only 72 individuals.

\begin{figure}[H]
  \begin{center}
  \vspace{-2cm} 
    \includegraphics[scale=0.5]{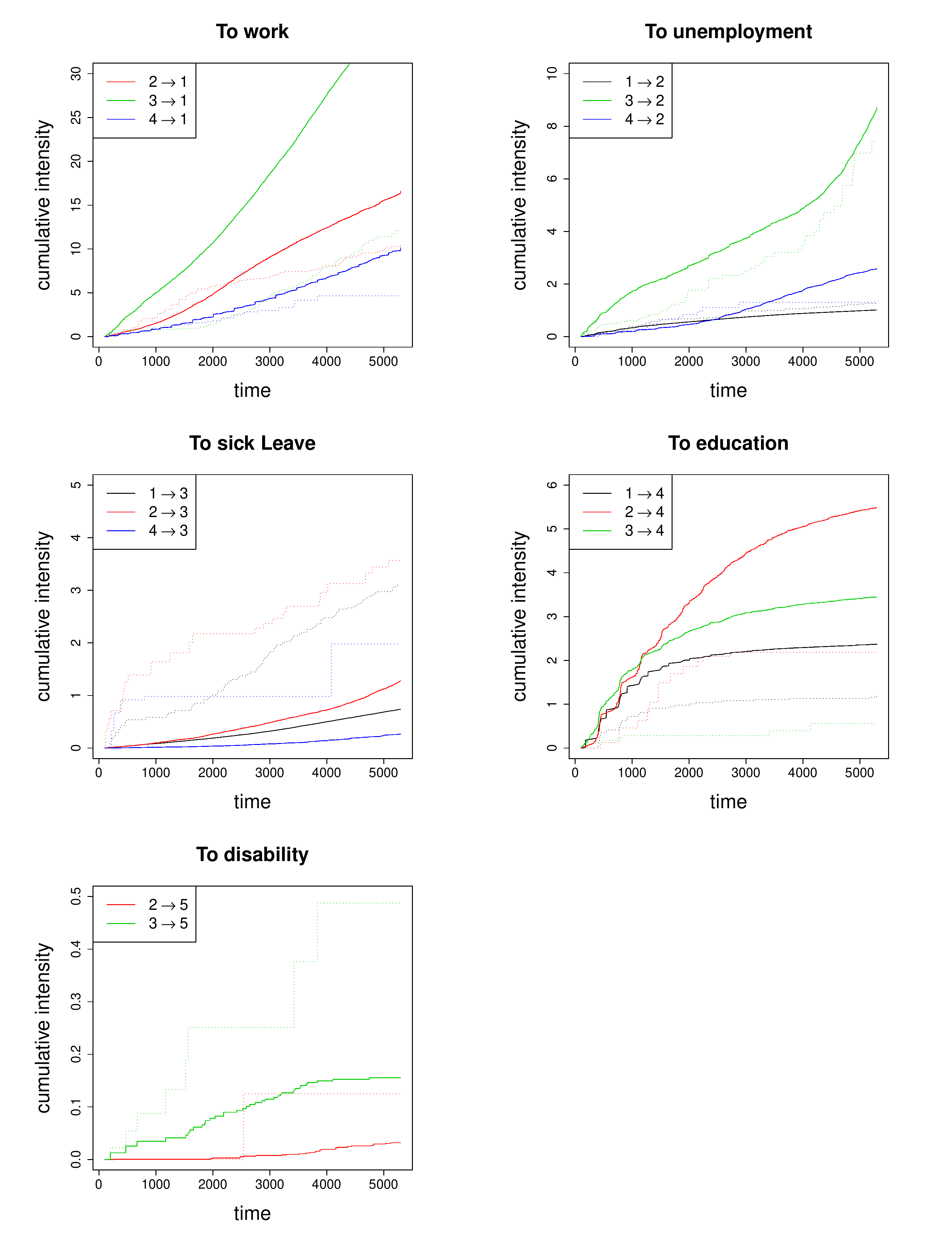} 
    \caption{Cumulative transition intensities starting at landmark time-point $s = 100$ days. Full drawn lines are Nelson-Aalen estimates based on the reduced cohort ($n=23288$), while dotted lines are Nelson-Aalen estimates based on the landmark sample of individuals in sick leave ($n=72$).}
    \label{cumhaz0} 
  \end{center}
\end{figure}

\begin{figure}[H]
  \begin{center}
    \includegraphics[scale=0.30]{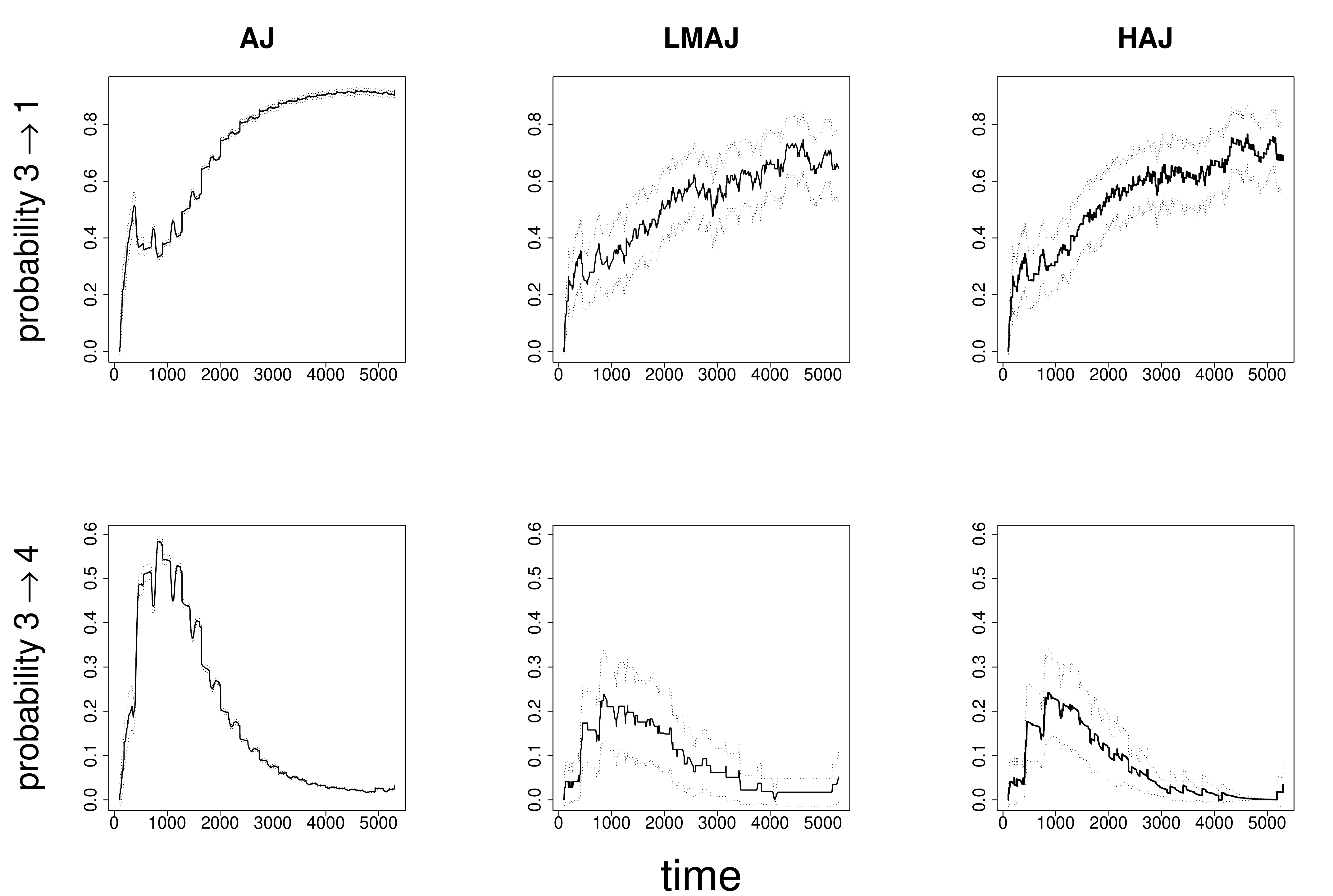} 
    \caption{Estimated transition probabilities from sick leave to work ($3 \to 1$) and from sick leave to education ($3 \to 4$). Dotted lines are 95$\%$ bootstrap (1000 samples) confidence intervals. The landmark sample consists of 72 individuals on sick leave at day 100. The full sample includes 23288 individuals.}
    \label{from3} 
  \end{center}
\end{figure}

We start by looking at cumulative transition intensities in the landmark sample and compare them with cumulative transition intensities calculated from the full dataset. The various intensities are shown in Figure \ref{cumhaz0}. Differences between the curves from the two data samples indicate a violation of the Markov assumption. Inspection of Figure \ref{cumhaz0} suggest that transitions exhibiting similar intensities in the landmark sample and the full sample are $1 \rightarrow 2$, $4 \rightarrow 1$ and $4 \rightarrow 2$. In addition to visual inspection we can test for non-Markov behaviour using the point-test described in Section \ref{sec3}.  Results from this test, found in Web Table C.1, imply that transitions $1 \rightarrow 2$, $2 \rightarrow 5$, $4 \rightarrow 1$ and $4 \rightarrow 2$ are not significantly non-Markov. 
Based on the above results, the HAJ estimator will here utilize all available data for transitions ($1 \rightarrow 2$, $2 \rightarrow 5$, $4 \rightarrow 1$ and $4 \rightarrow 2$) and only the landmark data for the other transitions. We illustrate the resulting transition probabilities from the landmark state (sick leave) into work and into education in Figure \ref{from3}. Estimated transition probabilities to all states based on the HAJ and LMAJ estimators are found in Web Appendix C.2.

In the current example, LMAJ and HAJ estimates are in close agreement and differ substantially from the AJ estimates. The AJ transition probability estimates are far greater than estimates produced by the LMAJ and HAJ estimators, indicating that the AJ estimates cannot be trusted. As for precision; Figure \ref{from3} indicates that the HAJ estimator has slightly higher precision than the LMAJ estimator. The AJ estimates naturally come with smaller confidence intervals due to the much larger sample size going into the estimation. Confidence intervals for the two landmark estimators are based on bootstrap, while for AJ, Greenwood plug-in estimates for standard errors were used. In Web Appendix C.2, we  also compare plug-in estimates of standard errors with bootstrap estimates for the HAJ estimator and find that they are very close, with the bootstrap standard errors being slightly larger.

\subsection{Example 2: Transitions from unemployment at $t = 3000$}

For the next example we still use data on high school completers of general education, but now consider individuals with cognitive scores between 4 and 6 (medium scores) and only individuals of parents who completed high school as their highest formal education. This amounts to a total of 10 451 individuals. Day 3000 is considered the landmark time point, and the landmark state is now unemployment. The landmark subset consists of 463 individuals. Results of log-rank tests for identifying Markov and non-Markov transitions can be found in Web Table C.2. The results indicate that transitions $2 \rightarrow 5$, $3 \rightarrow 2$, $3 \rightarrow 4$, $3 \rightarrow 5$, $4 \rightarrow 1$ and $4 \rightarrow 3$ are Markov. Thus, all available data are used for these transitions when constructing the hybrid estimator. 

\begin{figure}[H]
  \begin{center}
    \includegraphics[scale=0.30]{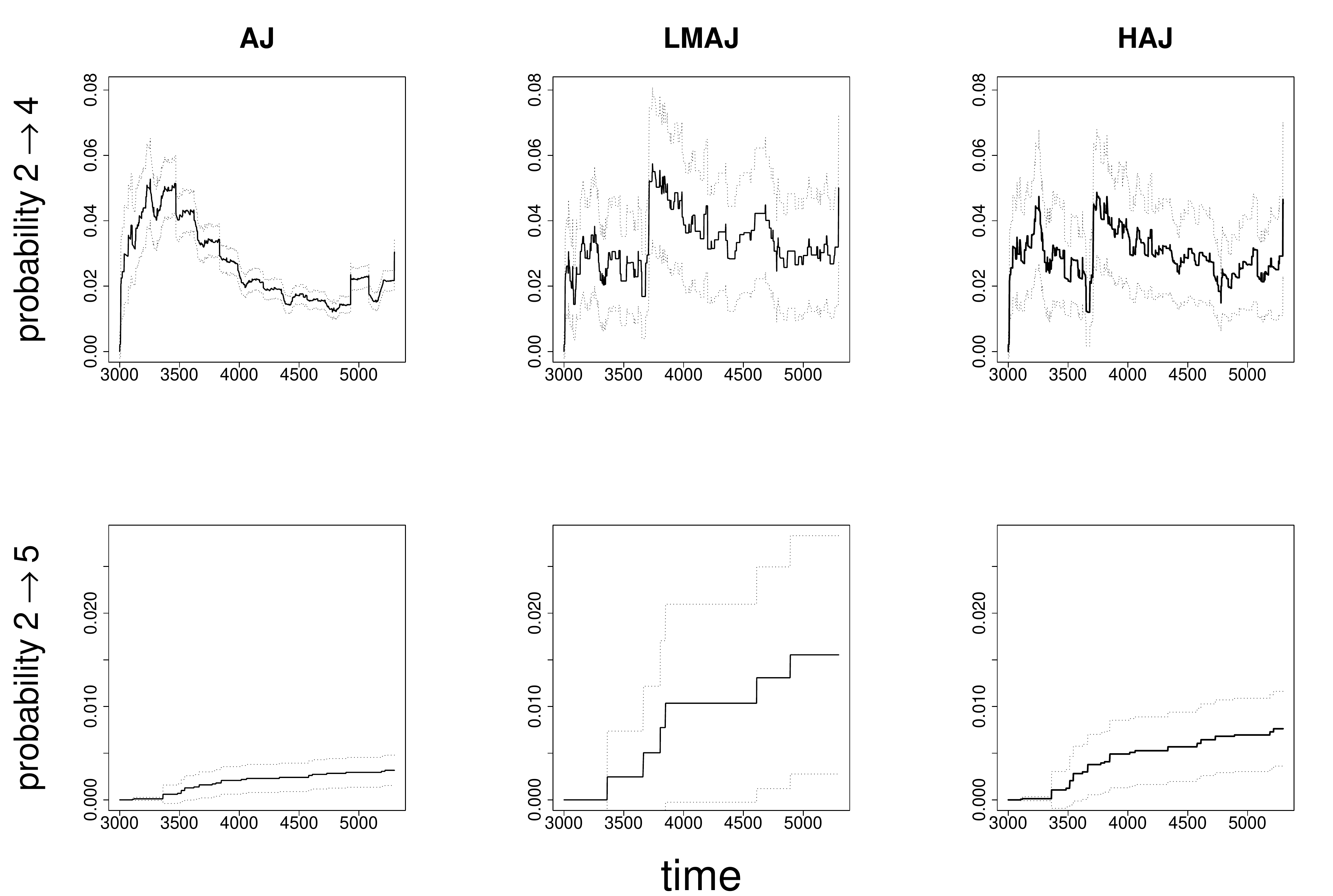} 
    \caption{Estimated transition probabilities from unemployment to education ($2 \to 4$) and from unemployment to disability ($2 \to 5$). Dotted lines are 95$\%$ confidence intervals: model based for AJ and bootstrap (1000 samples) based for HAJ and LMAJ. The landmark sample consists of 463 individuals unemployed at day 3000. The full sample includes 10451 individuals.}
    \label{from2} 
  \end{center}
\end{figure}

Estimates of transition probabilities from unemployment to education and disability are presented in Figure \ref{from2}. Compared to our previous example, AJ estimates are not as misleading when compared to the estimates from the two landmark methods, but do fail to capture the development in the first one third of the time period. In terms of precision, the HAJ estimator seem to give higher precision than the LMAJ estimator for both the showcased transitions, along with bootstrap and Greenwood type plug-in estimates of the standard errors.

\section{Discussion}
\label{sec6}

The idea behind the HAJ estimator is to utilize the interpretation of transition probabilities as a functional of transition specific rates and provide a framework for analyzing how specific transitions affect the estimation of transition probabilities. This sensitivity analysis point of view is useful when modelling non-Markov multi-state data. First, it frames the problem of non-Markov behaviour as a more familiar problem of bias-variance trade-off by considering the HAJ estimator as a compromise between the AJ (low variance) and the LMAJ (low bias) estimators. As a rule of thumb, one would generally expect the HAJ estimator to have higher bias (since one allows for Markov behaviour to be assumed in specific transitions) and lower variance (due to the increased sample sized) than the LMAJ estimator. The opposite, i.e. higher variance and lower bias, is generally expected when compared to the AJ estimator. Based on the simulation experiments it seems reasonable to believe that the lower variance comes at close to zero cost in bias. There are of cause exceptions to this rule and one should be aware that, depending on the data generating mechanism, the HAJ estimator may in principle be superior (in terms of bias or variance) to both or neither of the alternatives (AJ and LMAJ).  Secondly, with the HAJ estimator, one can think of the problem of non-Markov behaviour as a transition specific modelling choice, where certain transitions are more sensitive to non-Markov behaviour than others. Such considerations suggest a more comprehensive exploratory analysis of where, in a specific model, non-Markov behaviour can be problematic and where it might be negligible. In our construction of the HAJ estimator we focused on test statistics, but other tools such as judgement based on expert knowledge and visual inspection of plotted cumulative hazards or transition probabilities can also be considered as selection mechanisms. 

The choice between the HAJ, AJ and LMAJ estimator depends on what kind of non-Markov behaviour one is dealing with and how pronounced it is in the data. \cite{gunnes07} investigated the Datta-Satten estimator of state occupation probabilities in non-Markov models and reached to some extent a similar conclusion; the benefit of using an estimator which can handle non-Markov behaviour and is prone to bias under Markov regimes depends heavily on how much and why the model in question deviates from the Markov property. In large cohort studies one often has to assume that heterogeneity will be a problem simply due to the complexity of the underlying data generating mechanisms. Even in cases where the non-Markov behaviour is negligible this seems like a problematic assumption to start from. Furthermore, a trivial but important advantage of the HAJ estimator over the LMAJ estimator is the increase in the sample size. Besides a reduction in variance this might in practice mean a difference between a feasible and an infeasible estimator. The HAJ estimator is therefore relevant for many applications of non-Markov and partially non-Markov multi-state models, in particular for studies of limited sample size, where the LMAJ estimator is not a viable option.

\section*{Acknowledgements}

This work was funded by the Research Council of Norway (Grant No. 273674).
{\it Conflict of Interest}: None declared. \vspace*{-8pt}

\bibliographystyle{plainnat}  
\bibliography{references}

\section*{Supporting Information}
\label{sec10}

The \texttt{R} functions producing the simulation experiments (see \texttt{sim\_fun}), the point test and the grid test (see \texttt{wbMarkov}) are part of the \texttt{R}-package \texttt{multistate} available on github: \url{https://github.com/niklasmaltzahn/multistate}. The package \texttt{multistate} has a number of additional convenience functions useful for landmarking and hybrid estimation. Furthermore, the data output of \texttt{sim\_fun} is easily formatted for fitting multi-state models using the \texttt{mstate} package \citep{Putter11}.

Additional supporting information may be found online in the Supporting Information section at the end of the article. \vspace*{-8pt}



\section*{Supplementary material}

Supporting Information for *A hybrid landmark Aalen-Johansen estimator for transition probabilities in partially non-Markov multi-state models* by Maltzahn, Hoff, Aalen, Mehlum, Putter and Gran

August 25, 2020

# Web Appendix A

## On the estimation of variance of transition probabilities

Consider a number of independent individuals, $i = 1, \ldots, n$, and let $N_{jk}^{(i)}(t)$ be a counting process for transitions from $j$ to $k$ in the state space. The intensity process of individual $i$ relative to the complete history of the individual up to time $t$ is

$$\lambda_{jk}^{(i)}(t) = Y_j^{(i)}(t) Z_{jk}^{(i)}(t), \quad j, k \in E, \tag{A 1}$$

where $E$ is the state space and $Y_j^{(i)}(t)$ is 1 if the individual is in in state $i$ just prior to time $t$ and zero otherwise. The process $Z_{jk}^{(i)}(t)$ is the transition rate for individual $i$ in state $j$ at time $t$. These processes will generally differ between individuals according to the history of each individual. It is only when the process is Markovian that the transition rates will be fixed non-random quantities for all individuals.

According to counting process theory the processes

$$M_{jk}^{(i)}(t) = N_{jk}^{(i)}(t) - \int_0^t \lambda_{jk}^{(i)}(s) ds \tag{A 2}$$

are martingales.

Consider $n$ independent individuals. The history is now defined as the combined history of all individuals. The martingale property is then valid with respect to the combined history (due to independence). Hence, summing over individuals we get the martingales

$$\overline{M}_{jk}(t) = \overline{N}_{jk}(t) - \int_0^t \overline{\lambda}_{jk}(s) ds. \tag{A 3}$$

Define

$$\overline{Z}_{jk}(t) = \frac{1}{\overline{Y}_j(t)} \sum_i \left\{ Y_j^{(i)}(t) Z_{jk}^{(i)}(t) \right\}. \tag{A 4}$$

Then

$$\overline{M}_{jk}(t) = \int_0^t \left\{ d\overline{N}_{jk}(s) - \overline{Y}_j(s) \overline{Z}_{jk}(s) ds \right\} \tag{A 5}$$

is a martingale. Hence, the following expression is also a martingale:

1

$$\overline{M}_{jk}(t) = \int_0^t J_j(s) \left\{ \frac{d\overline{N}_{jk}(s)}{\overline{Y}_j(s)} - \overline{Z}_{jk}(s)ds \right\}, \tag{A 6}$$

where $J_j(s) = I(\overline{Y}_j(s) > 0)$ (see Aalen et al. ((2008, p. 88))).

The first part of the integral is the Nelson-Aalen estimator and it follows that it is an unbiased estimator for the quantity $\int_0^t J_j(s)\overline{Z}_{jk}(s)ds$, hence asymptotically unbiased for the expected value $\int_0^t E(\overline{Z}_{jk}(s))ds$.

From a standard martingale argument one may deduce the variance of the martingale as:

$$[\overline{M}_{jk}(t)] = \int_0^t J_j(s) \frac{d\overline{N}_{jk}(s)}{\overline{Y}_j(s)^2}. \tag{A 7}$$

However, this is not the variance of the Nelson-Aalen estimator in this case, due to the random quantities $\overline{Z}_{jk}(t)$. In order to deduce the full variance, these random quantities would need to be taken into consideration. In practice one would have to use bootstrap to handle this.

Hence, a tentative conclusion is that the standard variance for the Nelson-Aalen estimator is not valid. This implies that the plug-in variance estimator does not necessarily yield a correct variance for the LMAJ or HAJ estimator. However, it is unclear what the size of the bias is. The results in Section B of this supplementary material seem to indicate that the effect is minor, but more studies of this must be made. Sample size may be an important issue.

# Web Appendix B

# Supplementary material for the simulation study (Section 4 of the main document)

## B 1 Testing the Markov assumption

### B 1.1 Experiment 1

From Figure B 1 we see that the more right-skewed the frailty distribution described in Section 4 of the main document is, the larger is the rejection rates of the point-tests. We also see that the point-tests for transitions without frailties are close to the 5 percent level. The shaded areas in the test plots are 95% point-wise Agresti-Coull confidence intervals (see e.g. Agresti and Coull ((1998))). These confidence intervals have shown to exhibit robust small sample properties and end point properties (i.e. with success rates closed to zero or one). These are favourable features for the point-tests since the sample size of the tests shrinks with increasingly late landmark time points. Furthermore, the true rejection rates of Markov transitions should be closed to 0.05 (the $\alpha$ level) and end-point robustness is therefore also desirable. The same type of confidence intervals are produced for the grid-test.

**Rejection rates of point-tests**

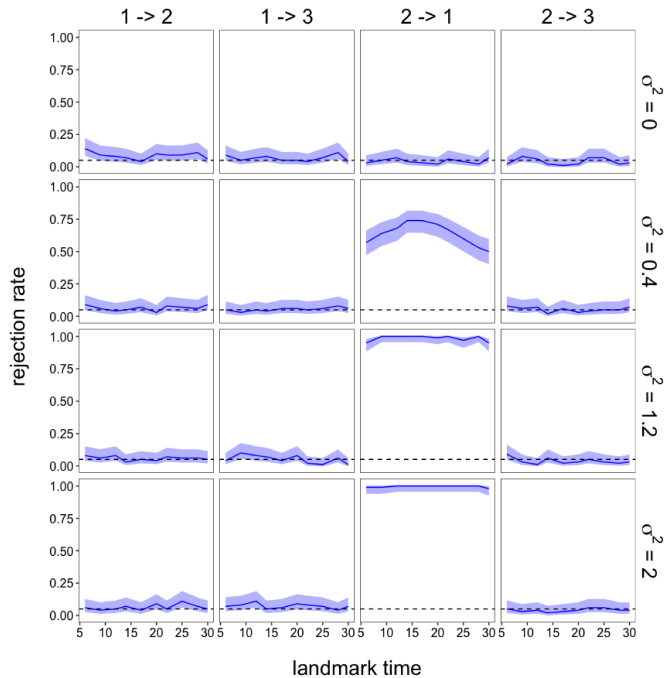

Figure B 1: Rejection rates of point-tests with Agresti-Coull confidence intervals plotted against landmark time. All numbers are based on 1000 samples where each sample has a size of 1000 individuals. The landmark grid is $\{6, 9, 12, 14, 17, 20, 22, 25, 28, 30\}$

The grid-tests in Table B 1 tell more or less the same story as the point-tests in Figure B 1, but are slightly more conservative. This behaviour is expected because of the construction being a supremum of many point-tests. Judging from the test statistics we see a clear indication of non-Markov behaviour in transition $2 \to 1$, but the main question is of course; is this behaviour

sufficiently non-Markov to suggest a change of estimator? To answer this question we can consider the MRSE plots.

**Grid-test rejection rates**

| Variance | Transitions | Mean | lower CI | upper CI |
|---|---|---|---|---|
| $\sigma^2 = 0.0$ | $1 \to 2$ | 0.13 | 0.08 | 0.21 |
|  | $1 \to 3$ | 0.07 | 0.03 | 0.14 |
|  | $2 \to 1$ | 0.08 | 0.04 | 0.15 |
|  | $2 \to 3$ | 0.05 | 0.02 | 0.11 |
| $\sigma^2 = 0.4$ | $1 \to 2$ | 0.03 | 0.01 | 0.09 |
|  | $1 \to 3$ | 0.04 | 0.01 | 0.10 |
|  | $2 \to 1$ | 0.98 | 0.93 | 1.00 |
|  | $2 \to 3$ | 0.02 | 0.00 | 0.07 |
| $\sigma^2 = 1.2$ | $1 \to 2$ | 0.08 | 0.04 | 0.15 |
|  | $1 \to 3$ | 0.10 | 0.05 | 0.18 |
|  | $2 \to 1$ | 1.00 | 0.96 | 1.01 |
|  | $2 \to 3$ | 0.05 | 0.02 | 0.11 |
| $\sigma^2 = 2.0$ | $1 \to 2$ | 0.05 | 0.02 | 0.11 |
|  | $1 \to 3$ | 0.03 | 0.01 | 0.09 |
|  | $2 \to 1$ | 1.00 | 0.96 | 1.01 |
|  | $2 \to 3$ | 0.03 | 0.01 | 0.09 |

Table B 1: Grid-test rejection rates for experiment 1, with upper and lower confidence bounds for each transition and each choice of frailty variance. All numbers are based on 1000 samples where each sample has a size of 1000 individuals. The landmark grid is $\{6, 9, 12, 14, 17, 20, 22, 25, 28, 30\}$

### B 1.2 Experiment 2

When comparing the plots in Figure B 2 with the MRSE plots for the transition probabilities in the main text, an interesting observation stands out. Looking at transitions $1 \to 2$ and $2 \to 3$, we see that whether or not we detect a large difference in the tests statistics for a particular transition, does not necessarily imply large differences for the corresponding transition probability. One reason for this is that the transition probability really depends on the relation between multiple hazards rather than one particular hazard function and the net effect on the transition probability of heterogeneity in multiple transitions is not clear. Furthermore, from the model specification, the test procedure has some natural constraints in terms of which groups it is able to compare based on landmark time and state. In other words the groups we are able to compare based on land-marking are not necessarily the ones who will reveal the underlying heterogeneous effects. Therefore we can have situations where the test will not be able to detect significant differences although they exist.

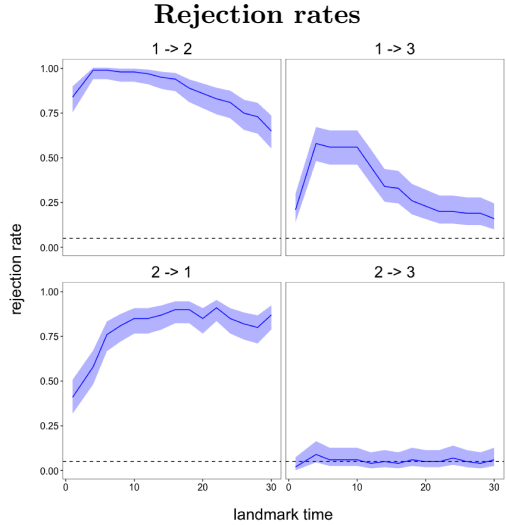

Figure B 2: Rejection rates for point-tests with confidence intervals. The landmark grid is $s = 1, 4, 6, 8, 10, 12, 14, 16, 18, 20, 22, 24, 26, 28, 30$.

|  | Mean | Lower CI | Upper CI |
|---|---|---|---|
| Transition $1 \to 2$ | 1.00 | 0.96 | 1.01 |
| Transition $1 \to 3$ | 0.69 | 0.59 | 0.77 |
| Transition $2 \to 1$ | 1.00 | 0.96 | 1.01 |
| Transition $2 \to 3$ | 0.03 | 0.01 | 0.09 |

Table B 2: Grid-test rejection rates for experiment 2 with upper and lower confidence bounds for each transition and each choice of frailty variance. All numbers are based on 1000 samples where each sample has a size of 1000 individuals. The landmark grid is $\{1, 4, 6, 8, 10, 12, 14, 16, 18, 20, 22, 24, 26, 28, 30\}$.

## B 2  Coverage and confidence intervals

To study the coverage and behaviour of Greenwood based confidence intervals we illustrate these for simulation experiment 1 and $\sigma^2 = 1.2$. See Figure B 3 and Figure B 4.

**Empirical coverage of point-wise confidence intervals based on the Greenwood type estimates for $\sigma^2 = 1.2$**

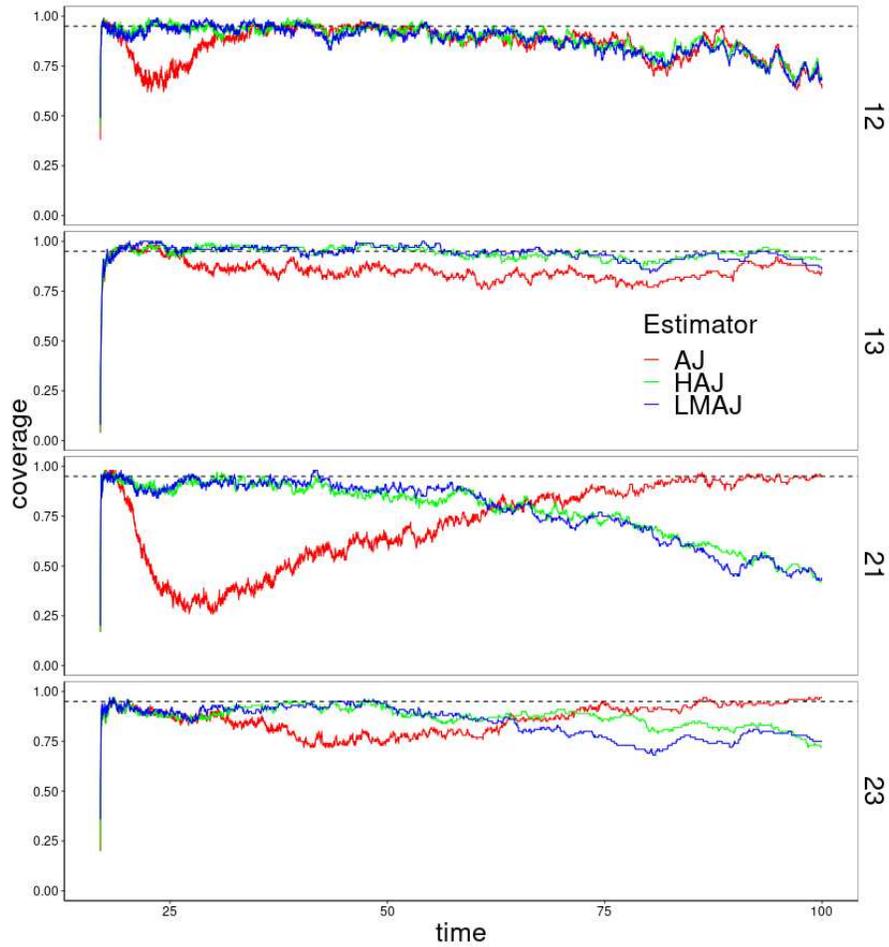

Figure B 3: A plot of the empirical coverage of the Greenwood type estimator for the standard error for each of the Estimators: AJ, HAJ and LMAJ. Estimates are based on 100 simulations of model 1 with $\sigma^2 = 1.2$. All estimates are computed from landmark time $s = 17$. The true transition probability is the mean of LMAJ estimates based on 1000 simulations with 1000 individuals in each.

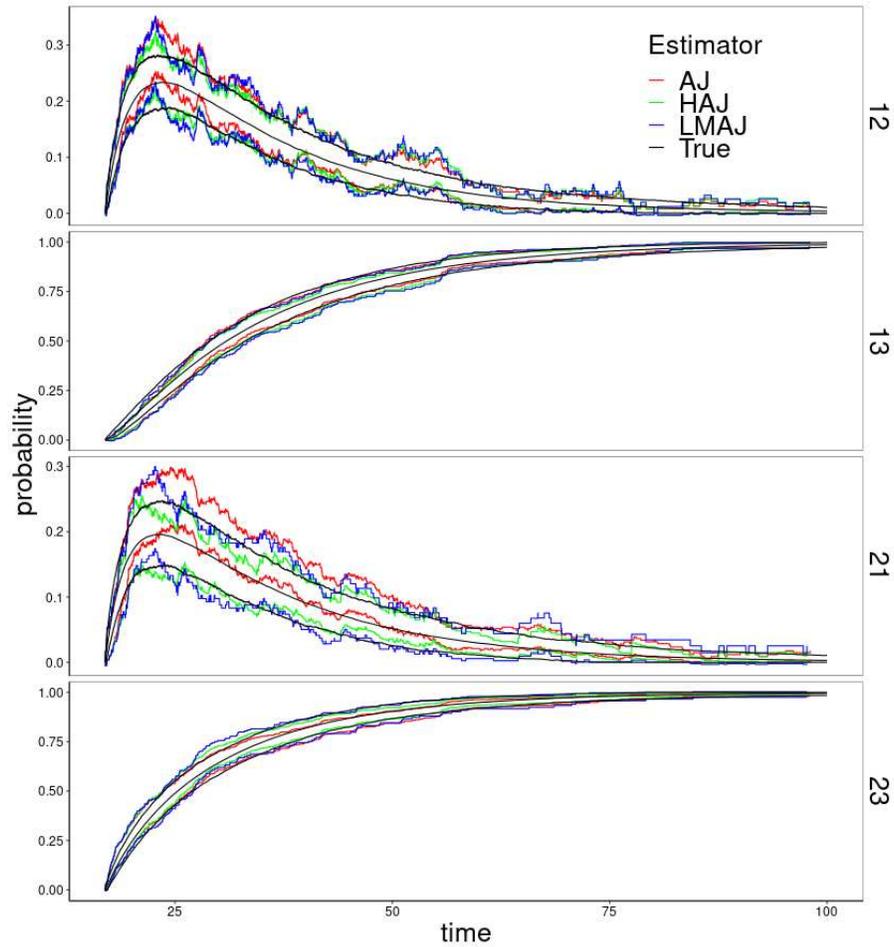

Figure B 4: A plot of pointwise confidence intervals of transition probabilities based on the Greenwood type estimator of the standard error. We see estimates of AJ, HAJ and LMAJ based on one simulations of model 1 plotted against the true confidence intervals for $\sigma^2 = 1.2$ and landmark time $s = 17$. The true confidence intervals are given by the 2.5 percentile and 97.5 percentile of LMAJ estimates based on 1000 simulations with 1000 individuals in each. The black line in the middle of the true confidence intervals are the true transition probabilities, i.e, the mean LMAJ estimate based on the 1000 simulations.

# Web Appendix C

# Supplementary material to the application on Norwegian registry data on sick leave, disability and work participation (Section 5 of the main document)

## C 1  Testing the Markov assumption

A formal test for identifying Markov and non-Markov transitions can be based on the log-rank test as described in Section 3.2 of the main document, testing for differences in transition intensities in the landmark sample and the rest of the data (full dataset minus the landmark subset). Results for the two examples in Section 5 of the main article follows in Table C 1 and Table C 2. Contrary to the simulation experiment, the p-values are here based on standard log-rank asymptotics; i.e. based on the chi-squared distribution and not on wild-bootstrapping.

| Transition | p-value | Transition | p-value |
|---|---|---|---|
| $2 \to 1$ | 0.000 | $2 \to 3$ | 0.000 |
| $3 \to 1$ | 0.000 | $4 \to 3$ | 0.000 |
| $4 \to 1$ | 0.067 | $1 \to 4$ | 0.000 |
| $1 \to 2$ | 0.206 | $2 \to 4$ | 0.000 |
| $3 \to 2$ | 0.000 | $3 \to 4$ | 0.000 |
| $4 \to 2$ | 0.255 | $2 \to 5$ | 0.134 |
| $1 \to 3$ | 0.000 | $3 \to 5$ | 0.000 |

Table C 1: Results from log-rank tests of differences in transition intensities in the landmark subset and the rest of the data, testing the Markov assumption for Example 1 (Section 5.1). The landmark subset is made up by all individuals who were on sick leave at day 100.

| Transition | p-value | Transition | p-value |
|---|---|---|---|
| $2 \to 1$ | 0.000 | $2 \to 3$ | 0.000 |
| $3 \to 1$ | 0.000 | $4 \to 3$ | 0.835 |
| $4 \to 1$ | 0.087 | $1 \to 4$ | 0.000 |
| $1 \to 2$ | 0.000 | $2 \to 4$ | 0.037 |
| $3 \to 2$ | 0.138 | $3 \to 4$ | 0.343 |
| $4 \to 2$ | 0.000 | $2 \to 5$ | 0.196 |
| $1 \to 3$ | 0.000 | $3 \to 5$ | 0.108 |

Table C 2: Results from log-rank test on the difference in transition intensities in the landmark sample compared to the rest of the data, testing the Markov assumption for Example 2 (Section 5.2). The landmark sample is made up by all individuals in unemployment at day 3000.

## C 2  Estimated transition probabilities and bootstrapped confidence intervals

To construct bootstrap confidence intervals for transition probabilities based on the HAJ estimator, we created 1000 bootstrap samples of the full dataset. From a full bootstrap sample we first create a bootstrap landmark sample. The bootstrap hybrid sample is then obtained by merging non-Markov transitions in the landmark sample with Markov transitions in the full bootstrap sample. The set of transitions deemed as Markov and non-Markov were assumed known in each

bootstrap iteration, and set to that resulting from the formal test on the original non-bootstrap dataset. For every bootstrap hybrid dataset, the transition probabilities were estimated from which we calculated the empirical standard errors.

## C 2.1 Estimated transition probabilities for Example 1 (Section 5.1)

See Figure C 1 for estimated transition probabilities from state 3 (sick leave) and Figure C 2 for a comparison of the corresponding estimated bootstrap and Greenwood type standard errors.

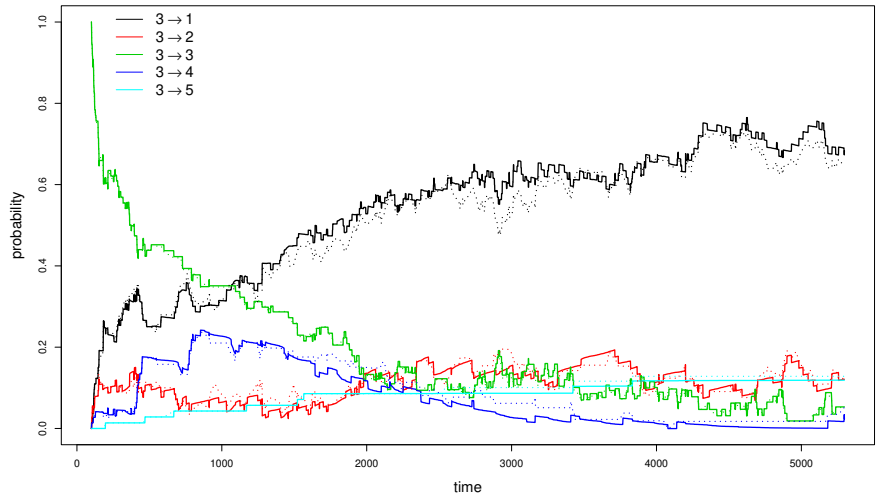

Figure C 1: Estimated transition probabilities from in state 3 (sick leave) at time $t = 100$. Full drawn lines are estimates from the HAJ estimator, dotted lines are from the LMAJ estimator.

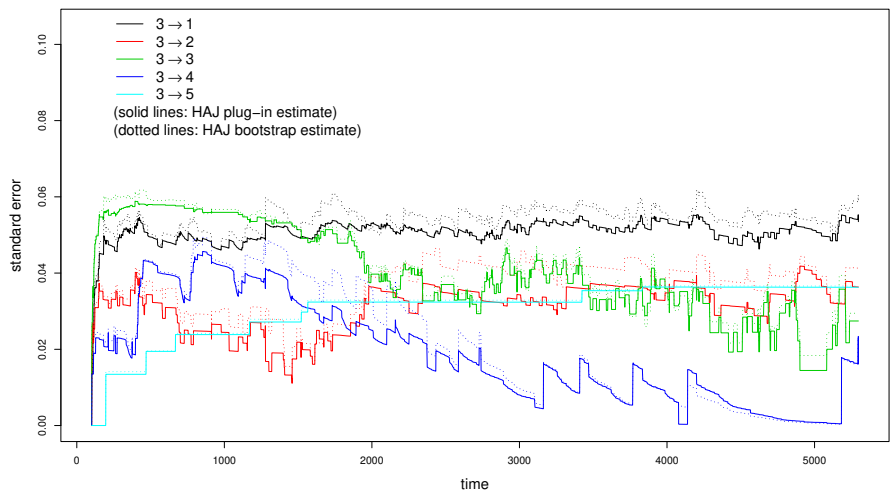

Figure C 2: Bootstrap (using 1000 bootstrap samples) and Greenwood type standard errors for HAJ estimates of transition probabilities from state 3 (sick leave) at day 100.

## C 2.2 Estimated transition probabilities for Example 2 (Section 5.2)

See Figure C 3 for estimated transition probabilities from state 2 (unemployment) and Figure C 4 for a comparison of the corresponding estimated bootstrap and Greenwood type standard errors.

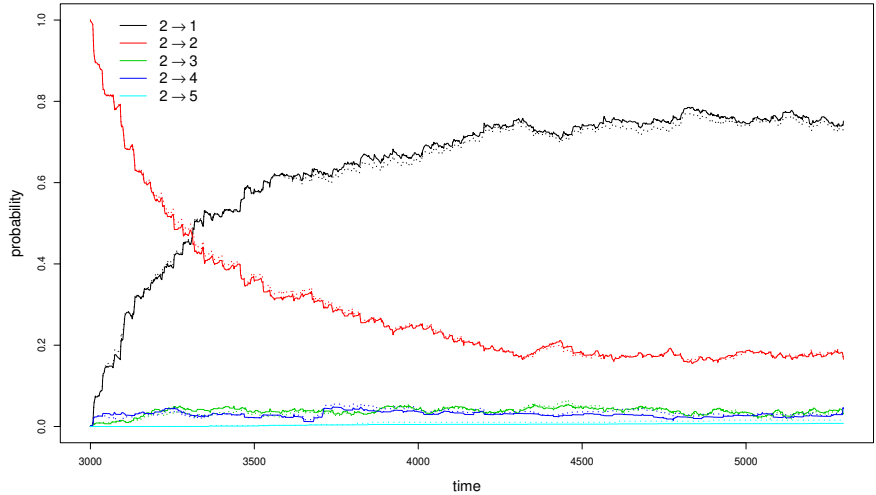

Figure C 3: Estimated transition probabilities from in state 2 (unemployment) at day 3000. Full drawn lines are estimates from the HAJ estimator, dotted lines are from the LMAJ estimator.

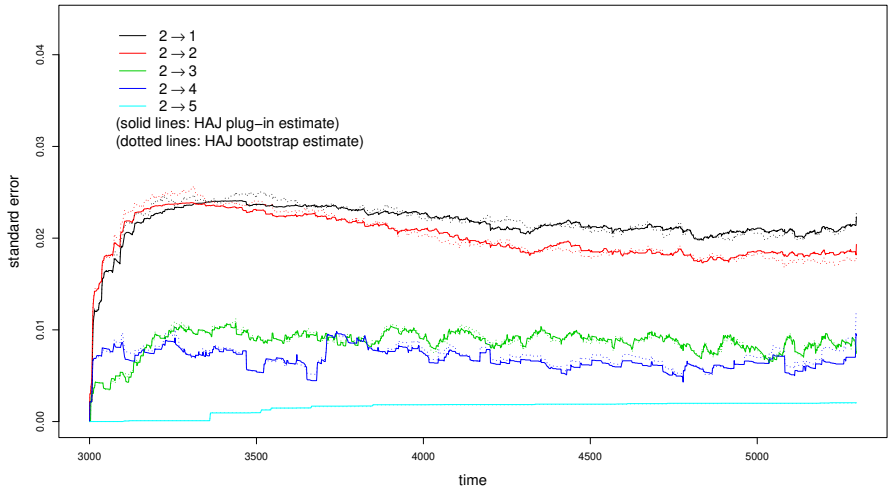

Figure C 4: Bootstrap (using 1000 bootstrap samples) and Greenwood type standard errors for HAJ estimates of transition probabilities from state 2 (unemployment) at day 3000.

# Web Appendix D

# Testable transitions

When constructing the HAJ estimator using two sample tests one needs to decide which transitions are amenable for testing. We will divide the complete set of transitions in the multi-state model into three disjoint sets $E = \mathcal{M} \cup \mathcal{N} \cup \mathcal{O}$. Here $\mathcal{M}$ is the set of Markov transitions (i.e., the set of transitions for which Markovianity is assumed without testing), $\mathcal{N}$ is the set of non-Markov transitions (for these landmark estimates are used in the HAJ), and $\mathcal{O}$ is the set of transitions that are open for testing. Obviously, different choices of $\mathcal{M}$, $\mathcal{N}$ and $\mathcal{O}$ may lead to different estimators. Choosing $\mathcal{N} = \mathcal{O} = \emptyset$ leads to the AJ estimator, while the choice of $\mathcal{M} = \mathcal{O} = \emptyset$ leads to the LMAJ estimator. An interesting observation is that many models will have untestable transitions. Take e.g. the transition from healthy state to death in an illness death model without recovery. In fact, if we by design cannot test a transition it is because the transition hazard estimate will be the same regardless of whether we landmark or not. Thus non-Markov behaviour only matters in so far as it is detectable in testable transitions. If we let $\mathcal{G} := (G, E)$ be the (bi-)directed graph representing transitions of $X$. A necessary and sufficient criteria for testing the transition $i \to j$ is the existence of subsets $U, V$ in $E$ satisfying:

$$U \cap V = \emptyset \text{ and there exists paths } p_U, p_V \text{ with root notes in } U,$$
$$\text{respectively } V \text{ such that } p_U \cap p_V = (i \to j). \tag{D 1}$$

From the criteria (D 1) we also see that if there exist a loop from $U$ to $U$, the transitions out of $U$ can be tested, at least in principle.



\end{document}